\documentclass[preprint,3p,times,12pt]{elsarticle}
\usepackage{graphicx}
\usepackage{dcolumn}
\usepackage{bm}
\usepackage{mathtools}
\usepackage[dvipsnames]{xcolor}
\usepackage{caption}
\usepackage{subcaption}
\usepackage{amsfonts}
\usepackage{setspace}
\usepackage[T1]{fontenc}
\usepackage{lipsum}
\journal{13th USNCM}
\date{January 18, 2023}

\begin{document}

\begin{center}
$13^{th}$ U.S. National Combustion Meeting \\ Organized by the Central States Section of the Combustion Institute \\ March 19-23, 2023 \\ College Station, Texas
\end{center}
\begin{frontmatter}
\title{A Framework for Combustion Chemistry Acceleration with DeepONets}


\author[NCSU]{Anuj Kumar}
\author[NCSU,*]{Tarek Echekki}
\affiliation[NCSU]{organization={Department of Mechanical and Aerospace Engineering, North Carolina State University},
            addressline={Campus Box 7910}, 
            city={Raleigh},
            postcode={27695}, 
            state={NC},
            country={USA}}
\affiliation[*]{Corresponding Author Email: techekk@ncsu.edu}
\begin{abstract}
\footnotesize
A combustion chemistry acceleration scheme is developed based on deep operator nets (DeepONets). The scheme is based on the identification of combustion reaction dynamics through a modified DeepOnet architecture such that the solutions of thermochemical scalars are projected to new solutions in small and flexible time increments. The approach is designed to efficiently implement chemistry acceleration without the need for computationally expensive integration of stiff chemistry. An additional framework of latent-space dynamics identification with modified DeepOnet is also proposed which enhances the computational efficiency and widens the applicability of the proposed scheme. The scheme is demonstrated on simple chemical kinetics of hydrogen oxidation to more complex chemical kinetics of n-dodecane high- and low-temperature oxidations. The proposed framework accurately learns the chemical kinetics and efficiently reproduces species and temperature temporal profiles corresponding to each application. In addition, a very large speed-up with a great extrapolation capability is also observed with the proposed scheme.
\end{abstract}

\end{frontmatter}

\section{Introduction}
Chemistry integration is the principal bottleneck in combustion simulations. Chemistry reduction if implemented correctly can significantly decrease the stiffness of chemistry mechanisms. In addition to chemistry reduction, different strategies for chemistry acceleration have been proposed, including chemistry tabulation (see for example the use of {\it {in situ}} adaptive tabulation (ISAT)~\cite{Pope1997}), the piecewise reusable implementation of solution mapping (PRISM)~\cite{Tonse2003}, adaptive chemistry, including dynamic approaches \cite{liang2009}, \cite{Continuo2011}, \cite{Sun2017} and \cite{DAlessio2020}, manifold-based methods, such as intrinsic low-dimensional manifolds (ILDM)~\cite{maas1992} and computational singular perturbation (CSP) methods~\cite{Lam1994}. Chemistry acceleration is often combined with operator splitting to decouple chemistry integration from the remaining transport terms in reacting flows. Artificial neural networks (ANNs) have also been used to develop regressions for chemical source terms \cite{Christo1996a,Christo1996b,Blasco1998,Blasco2000, Ranade2019c, Ranade2019d, Chen2000, Wan2020, Wan2021, Alqahtani2021,Echekki2023}).

In recent years, several approaches were proposed to accelerate chemistry integration in combustion beside the implementation of machine learning (ML) as tools for chemistry regression. They include the ChemNODE approach proposed by Owoyele and Pal~\cite{ChemNODE} and the methods proposed by Zhang et al.~\cite{Zhang2021AIAA} and Galassi et al.~\cite{Galassi2022}. ChemNODE combines a stiff ODE solver and a deep neural networks that minimizes the solution of chemistry integration based on the law of mass action and the ANN-based reaction rates. Zhang et al.~\cite{Zhang2021AIAA} trained a deep neural network (DNN) to project a solution of the thermo-chemical state vector between two time steps. Galassi et al.~\cite{Galassi2022} proposed the use of ANN regression as a cheaper surrogate to the local projection basis within the CSP approach~\cite{Lam1994}, thus lessening the bottleneck within the CSP chemistry acceleration technique.

Adopting an effective ML-based chemistry integration method must address the following constraints. First, ML tools rely on the ability of complex and potentially deep neural networks to reduce uncertainties in the selection and predictions of inputs and outputs, respectively, in these networks. Given the inherent dimension of chemical systems, this represents a compounded challenge to the nonlinear relations between inputs and outputs. It is often easier to demonstrate a chemistry integration technique for simpler fuels; yet, the real challenges arise with more complex fuels. Second, chemistry integration can be achieved by either reducing the dimensionality describing the chemical system and implementing computationally-efficient evaluations of reactions or projection schemes. Projection methods that predict the solution vector at time increments from a current time step remain an attractive choice as they avoid the direct integration of stiff chemistry. Such methods have been shown to be very efficient, such as in the case of the ISAT approach~\cite{Pope1997}. Third, selecting the appropriate data for training the acceleration method is crucial to avoid the inherent cost of {\it{in situ}} training within a simulation.

Among the ML-based complex networks proposed in recent years, deep operator networks, or DeepONets, \cite{deeponet_orig} have emerged as effective networks to learn continuous nonlinear operators mapping functions between infinite-dimensional spaces based on the universal approximation theorem for operators \cite{universal_operator}. Utilizing that, Wang et. al. \cite{longtime} proposed a framework for long-time integration of time-evolving ODEs which projects solution vectors of an ODE in small time increments. The framework produces stable results along a long-time horizon; however, the network size grows rapidly with the solution vector dimensions. With DeepONets, it is important to manage the size of the network for its effective implementation in combustion chemistry acceleration.

The objective of this study is to develop an efficient combustion chemistry acceleration based on DeepONet. The method is based on the projection of the solution of thermochemical scalars' vector into a new solution at a small and flexible time increment similar to the approach of Wang et. al. \cite{longtime}. Strategies to enhance computational efficiency and widen the applicability of the framework to a combustion reaction system are proposed.  Additionally, a method to advance the solution vector in the reduced space is developed.  The proposed framework eliminates the need for computationally expensive stiff chemistry integration. The framework is validated using two hydrogen and n-dodecane low- and high-temperature chemistry oxidations.

\section{Methodology} \label{}
In this section, a summary of the DeepONet framework is presented followed by the implementation of DeepONet for chemistry acceleration.

\subsection{Deep Operator Net (DeepOnet)}
The DeepOnet aims to learn an operator, $G$, which takes an input function $u$ and gives an output function $G(u)$. The output function, $G(u)$, is evaluated at any time instant or location, $y$, and results in a real number, $G(u)(y)$.

The network input consists of two separate components $\left[u\left(x_1\right), u\left(x_2\right), \ldots, u\left(x_m\right)\right]^T$ and $y$ and both the inputs are handled separately by two neural networks.
 The first sub-network is the trunk net, which takes the independent variable, $y$, as an input and outputs $ [t_1, t_2, \ldots,t_p ]^T \in \mathbb{R}^p $.
 The second sub-network takes input $\left[u\left(x_1\right), u\left(x_2\right), \ldots, u\left(x_m\right)\right]^T$ at fixed sensor points, $m$, and outputs $\left[b_1, b_2, \ldots, b_p\right]^T \in \mathbb{R}^p$. The final output of the DeepOnet is obtained by merging the outputs of trunk and branch network via a dot product:
 \begin{equation}\label{eq:general}
 G(u)(y) \approx \sum_{k=1}^p \underbrace{b_k\left(u\left(x_1\right), u\left(x_2\right), \ldots, u\left(x_m\right)\right)}_{\text {branch }} \underbrace{t_k(y)}_{\text {trunk }}
 \end{equation}

The DeepOnet guarantees the universal approximation of an operator by its construction and it has been shown to accurately approximate various implicit and explicit operators \cite{deeponet_orig, Mao2021, WangWangPerdikaris2021}.

\subsection{Integration framework for chemical kinetics: React-DeepOnet}
The chemical kinetics of any fuel is described by a system of ODEs, which are represented as an initial value problem taking the form
\begin{equation}
    \frac{d\phi}{dt} = S(\phi, p_0) \quad  t \in[0, T]
    \label{eqn1}
\end{equation}
 \begin{equation} 
    \phi( 0)= \phi _0
\end{equation}
where $\phi$ is the thermochemical composition vector, which includes species mass fractions, $Y_{i}$, and Temperature. The chemical source term, $S$ is a function of thermochemical scalars, $\phi_0$, and represents the chemical reaction source terms based on the law of mass action. The reaction system is stiff and computationally expensive to solve since all the reactions temporally evolve on different time scales and involve nonlinear evaluations of multiple chemical reactions.

The solution profiles for a reaction system uniquely depend on the initial conditions, $\phi_0$. A DeepOnet can provide a solution of the reaction system (see Eq.~\ref{eqn1}), $\phi(t)$, at any time, $t$, by operating on the initial condition, $\phi(0)$. In this framework, 
Branch net and Trunk net take $\phi(0)$ and $t$ as inputs respectively and the output of the DeepOnet is the solution, $\phi(t)$ \cite{deeponet_orig, WangWangPerdikaris2021}. However, with this strategy, predictions at very large times may be inaccurate and unstable \cite{longtime, Venturi2023}. In addition, a huge training data set is needed in order to have low generalization errors. 

\begin{figure}[h!]
\centering 

\begin{subfigure}{0.55\textwidth}
\centering
\includegraphics[width=\textwidth]{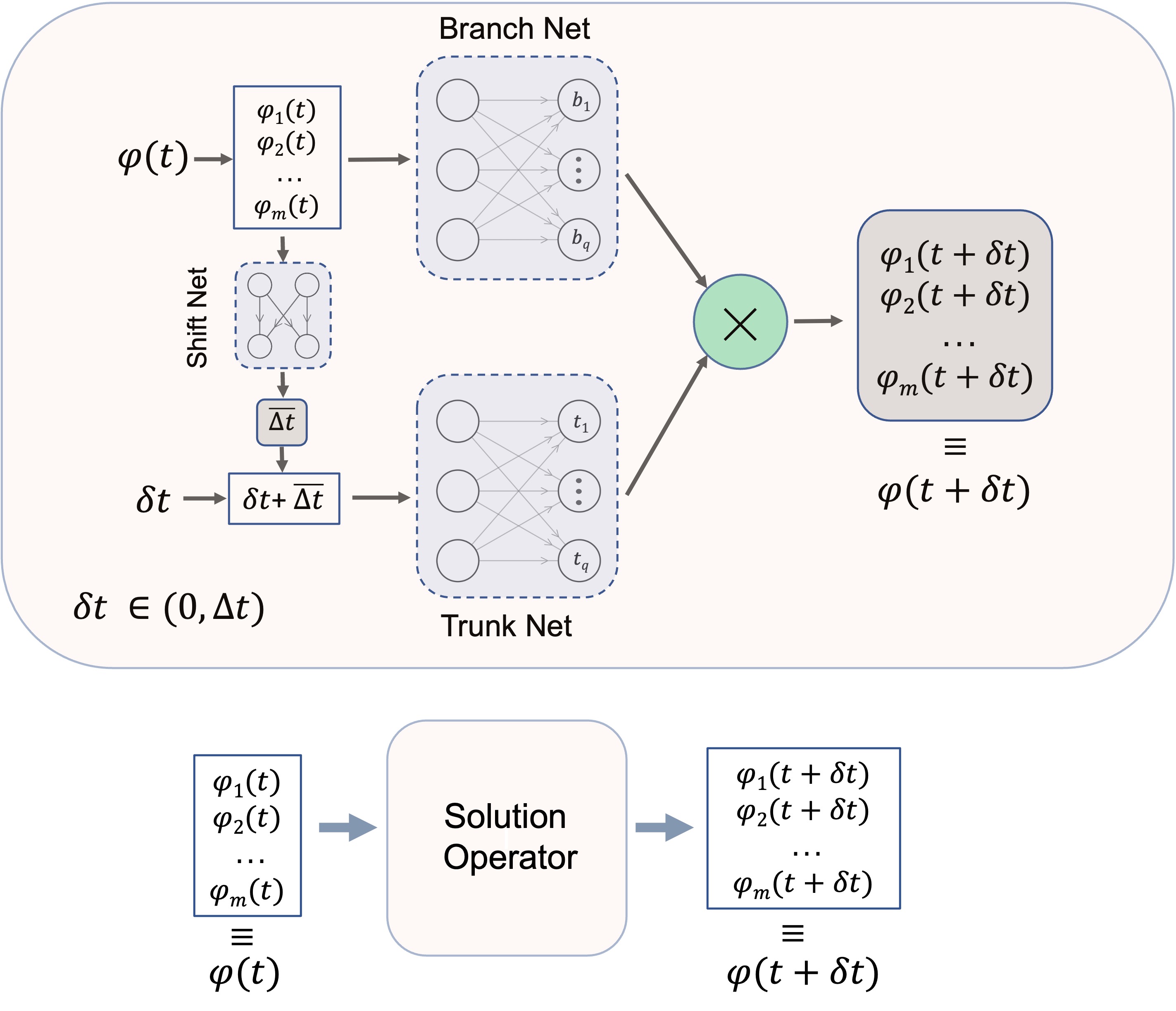} 
\caption{Time integration framework for chemical kinetics: React-DeepOnet}
\label{longtimeintegration}

\end{subfigure}
\hfill
\begin{subfigure}{0.35\textwidth}
\centering
\includegraphics[width=\textwidth]{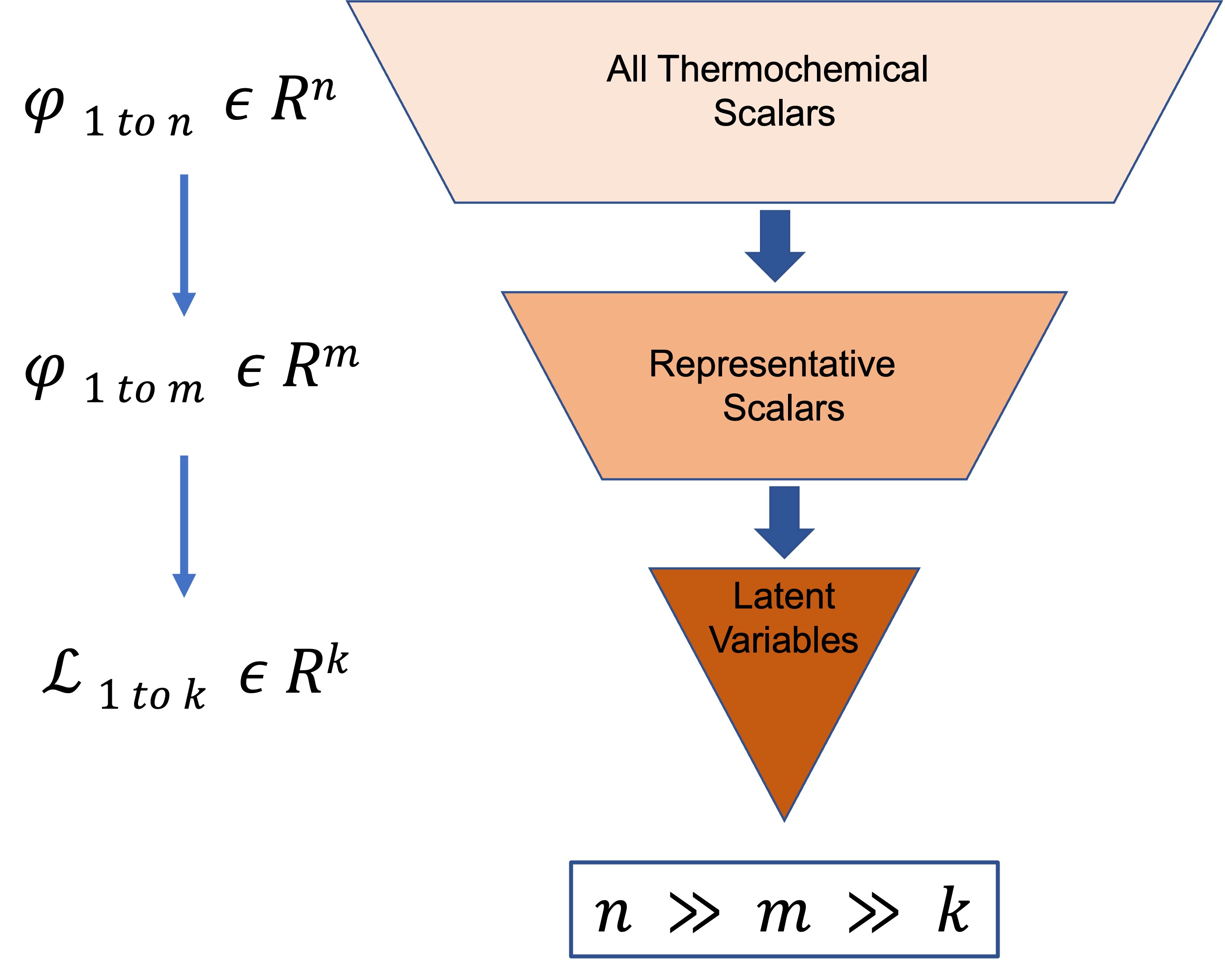} 
\caption{Thermochemical scalar hierarchy}
\label{scalarhierarchy}
\end{subfigure}

\end{figure}

Instead, we learn the solution operator and find the solution profile in a short period of $t \in[0, \Delta t]$ for a particular initial value of the thermochemical vector. After that, the solution values obtained from DeepOnet at $t= \Delta t$, $\phi(\Delta t)$, are fed into the DeepOnet as an initial condition, and the solution profile in the next window $t \in[\Delta t, 2\Delta t]$ is obtained. Continuing this iteratively, a whole solution profile for the reaction system in  $t \in[0, T]$ is acquired. The strategy is adopted from Wang et. al. \cite{longtime} where a physics-informed DeepOnet is deployed for time-evolving ODEs. The strategy is depicted in Figure- ~\ref{longtimeintegration}.

In Figure- ~ \ref{longtimeintegration}, apart from Branch net and Trunk net, another neural network, termed as ShiftNet, is added to DeepOnet. It takes the same input as Branch net and outputs a real number, which augments the inputs of the Trunk net.
The ShiftNet is inspired by the work of Venturi et. al. \cite{Venturi2023} and Hadorn et. al. \cite{hadorn2022shift} and provides an optimum scaling for inputs of the Trunk net. Strong parallelism between low-order approximation with SVD and operator approximation with DeepOnets has been shown \cite{Venturi2023, Liu2022DeepPropNetA}. SVD rank explodes and low-order approximation fails when transitional symmetry is present in the input data columns. With proper scaling and centering of the input data, a few leading singular modes can capture the whole behavior. The solution profiles for different initial conditions for a particular combustion reaction system are similar in nature and overlap with almost a single profile with proper scaling. The addition of the ShiftNet into the DeepOnet allows the automatic discovery of an optimum scaling to have a smaller DeepOnet architecture for a given prediction accuracy. 
This framework of operator leaning over a small time window and optimal scaling of similar solution profiles with ShiftNet in DeepOnet is termed as React-DeepOnet.

\subsection{Training Data Generation}

Solution profiles for each reaction system with different initial temperatures and equivalence ratios are obtained from Cantera \cite{cantera} with the Ideal Gas Constant Pressure 0D homogeneous reactor. For each fuel, the reaction is evolved until equilibrium is reached. The solution vector comprising reactants, products, and temperature is stored at a fixed time interval in each case.
Figure- ~\ref{scalarhierarchy} shows representative scalars are the subset of the full thermochemical scalars and comprise important reactants, products, and major intermediates. React-DeepOnet is used to evolve only a representative set of species, chosen to identify combustion dynamics and correlate with the remaining scalars. The remaining scalars can be reconstructed using an ANN \cite{Owoyele2017,mirgol2015}.
\[
[\phi_1, \phi_2, \dots, \phi_m] \in [\phi_1, \phi_2, \dots, \phi_n]
\]
 In order to keep track of those minute variations before the reaction, a $5^{th}$ root transformation is performed on the whole representative dataset and after that linearly scaled to [-1, 1].
\\
One input consists of two variables each for the Branch net and Trunk net and is paired with a real number as an output. The input-output pairs are extracted from a temporal window of $(0, \Delta t)$. This window of $(0, \Delta t)$ moves from initial time ($t=0$) to the final time  ($t=T$). Each point on the solution profile (before the last window on the profile) serves as an initial condition such that inputs are \{$\phi_0^{i}$, ($t_j$)$_{j=1}^{N_{win}}$ \}$_{i=1}^{N_{sol}}$ to Branch net and Trunk net and corresponding outputs are \{($\phi (t_j) ^{i}$)$_{j=1}^{N_{win}}$ \}$_{i=1}^{N_{sol}}$. Where, $N_{Sol}$ is the number of solution points in a single profile, $N_{Win}$ is the window size of $(0,\Delta t)$. This framework allows the generation of a relatively large data set with a limited number of 0D homogeneous reactor simulations from Cantera.

\subsection{Latent Space Dynamics Identification}

\begin{figure}[h!]
	\centering
		\includegraphics[scale=.08]{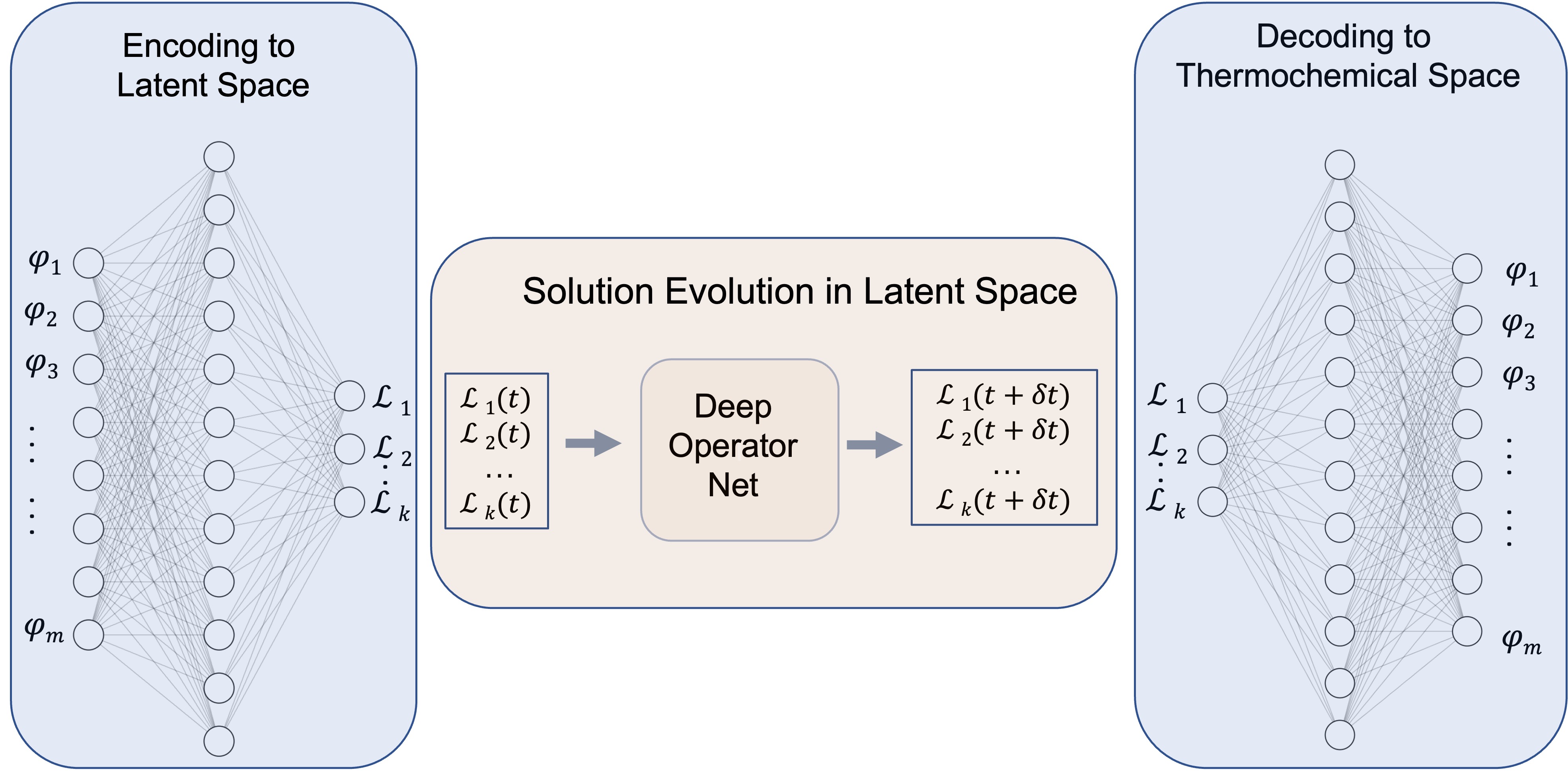}
	\caption{Latent Space Dynamics Identification Framework}
	\label{latentdynamics}
\end{figure}

The dimensionality of the representative dataset can be further reduced with various dimensionality reduction techniques without incurring any loss in its representation capability. For this purpose, an Autoencoder is deployed, which has been shown to effectively reduce the dimensionality and identify the latent space \cite{mirgol2015, brunton2019, carlberg2020, zhang2022}. Autoencoders provide a nonlinear mapping from a high dimensional space ($\mathbb{R}^m$) to a low dimensional space ($\mathbb{R}^k$). Figure- ~\ref{scalarhierarchy} depicts the order of dimensionality reduction starting from a full thermochemical space (which has hundreds of scalars) to a few latent variables.  

Once the representative scalars are condensed down to the latent variables, the dynamics are learned in that space \cite{brunton2019, zhang2022,grassi2022reducing,fries2022}. For this purpose, our framework of long-time integration with React-DeepOnet is deployed. The strategy is demonstrated in Figure-~\ref{latentdynamics} where the solution evolves in the reduced space. The thermochemical scalars can be reconstructed with the decoder once the desired system state is reached.

\section{Results} \label{results}

The long-time integration framework of React-DeepOnet is evaluated on the 0-dimensional H$_2$ and n-dodecane combustion. The solution profiles for H$_2$ are
generated at initial temperatures $T_i = $ 1020, 1065, 1110, 1155 and 1200 K and equivalence ratios $\Phi_i = $ 0.6, 0.7, 0.8, 0.9, 1.0, 1.1, 1.2 and 1.3 at a constant pressure of 1atm. On the other hand, two separate cases of high-temperature chemistry and low-temperature chemistry are considered for n-dodecane combustion. 11 solution profiles with initial temperatures equispaced between 1400-1500K at a constant pressure of 1 atm are considered for high-temperature n-dodecane combustion. Low-temperature n-dodecane solution profiles are generated at a constant pressure of 20 atm and with initial temperatures $T_i = $ 750, 760,770, 780, 790, 800, 810, 820, 830 and 840 K. Both n-dodecane combustion solution profiles are generated at a fixed equivalence ratio of $\Phi_i = $ 1.0.

Five representative thermochemical scalars, $T$, H$_2$, H, O$_2$ and H$_2$O, are selected out of the whole solution vector containing 11 thermochemical scalars ([T,H$_2$,H,O$_2$,O,OH,H$_2$O,HO$_2$, H$_2$O$_2$,AR,N$_2$ ]) for H$_2$ oxidation. While n-dodecane combustion reaction mechanism JetSurF2.0 \cite{wang2010high} consists of 349 thermochemical scalars (348 species and temperature). Out of these,  a subset of 5 representative scalars [$T$, CO, H$_2$O, CH$_3$ and C$_3$H$_6$]  is considered for high-temperature n-dodecane combustion. Low-temperature n-dodecane combustion has more complex chemical interactions and a bigger set of 8 representative scalars [$T$, CO, H$_2$O, H$_2$, H, O$_2$, C$_2$H$_4$, and C$_4$H$_{81}$] is identified. The reduction from the full set to the representative set is performed based on Alqahtani and Echekki~\cite{Alqahtani2021}. The same training dataset of low-temperature chemistry is also used for latent space React-DeepOnet. However, a larger representative set of 14 scalars is identified and reduced to 3 latent variables with an autoencoder of size \{14, 100, 20, 3, 20, 100, 14\}.

A time window of ($0, 20\mu s$) and ($0, 100\mu s$) is chosen to learn the solution operator with React-DeepOnet such that the maximum time-advancement in a single prediction is $\Delta t= 20 \mu s$ and $\Delta t= 100 \mu s$ for H$_2$ and n-dodecane combustion respectively. 
The React-DeepOnet model is tested on all the initial temperatures at equivalence ratios $\Phi = $ 0.6, 0.7, 1.0, 1.3 in case of H$_2$ combustion. On the other hand, high-temperature n-dodecane React-DeepOnet is tested on 1420, 1450, 1480 K initial temperatures while low-temperature n-dodecane is tested on initial temperatures of 780 K and 820 K.

The React-DeepOnet architecture is modeled and trained with JAX \cite{frosting2018} framework utilizing Just in Time compilation and Auto-vectorization on GPUs to significantly reduce the training time. The model is trained on an NVIDIA A100 GPU with Adam optimizer \cite{kingma2014}. The model parameters and hyperparameters for all the cases are listed in Table-~\ref{tab:table2}.

 \begin{table}[h!]
 \begin{center}
    \caption{React-DeepOnet Training parameters and Speed-Up}
    \scalebox{0.70} {
    \label{tab:table2}
    \begin{tabular}{c|c|c|c|c} 
      Parameters & H$_2$ & High-Temp & Low-Temp & Latent-Space \\
       &  & n-dodecane & n-dodecane & Dynamics \\
      \hline
       Prediction Window ($0-\Delta t$) & ($0- 20 \mu s $)  & ($0- 100 \mu s $) & ($0-100 \mu s $) & ($0- 100 \mu s $)\\
       Branch Net & [5,80,80,200]& [5,100,100,100,100,200] & [5,$\underbrace{ 100,\dots,100}_{\times 8}$,480] & [3,100,100,100,100,180]\\
       Trunk Net & [1,80,80,200] &[1,100,100,100,100,200] & [1,$\underbrace{ 100,\dots,100}_{\times 8}$,480]& [1,100,100,100,100,180] \\
      ShiftNet & [5,10,10,1]  & [5,20,20,1] & [8,40,40,1] & [3,20,20,1\\
      Total Parameters & 47 k  & 104 k & 244 k & 99 k \\
      Training Iterations & $2\times 10^5$  & $1\times 10^5$ & $1\times 10^5$ & $1.2\times 10^5$ \\
      Mini-Batch size  & 20,000  & 30,000 & 30,000 & 30,000 \\
     Learning Rate & $1\times 10^{-2}$ & $8\times 10^{-3}$ & $1\times 5^{-3}$ & $8\times 10^{-3}$ \\
       Training Time (min)  & 10.5  & 6.5 & 11.3 & 6.4 \\
       Final time on solution trajectory (T)(ms)  & 0.5  & 1.0 & 5.0 & 5.0 \\
       Integration time step ($\delta t$)  & 20$\mu s$  & 100$\mu s$ & 100$\mu s$ & 100$\mu s$ \\
       React-DeepOnet integration CPU time (ms)  &  1.86 & 0.70 & 4.48 & 2.5 \\
       Cantera Integration CPU time (ms)  & 12.4  & 8770 & 10500 & 10500 \\
       Speed-Up  & 6.7  & 12,500 & 2,340 & 4,200 \\
       Mean Absolute Error  & $0.1 \%$  & $0.2 \%$ & $0.6 \%$ & $0.5 \%$ \\
       \hline
    \end{tabular}
     }
\end{center}     
\end{table}

\begin{figure}[h!]
\centering 

\begin{subfigure}[h]{0.48\textwidth}
\includegraphics[width=\textwidth]{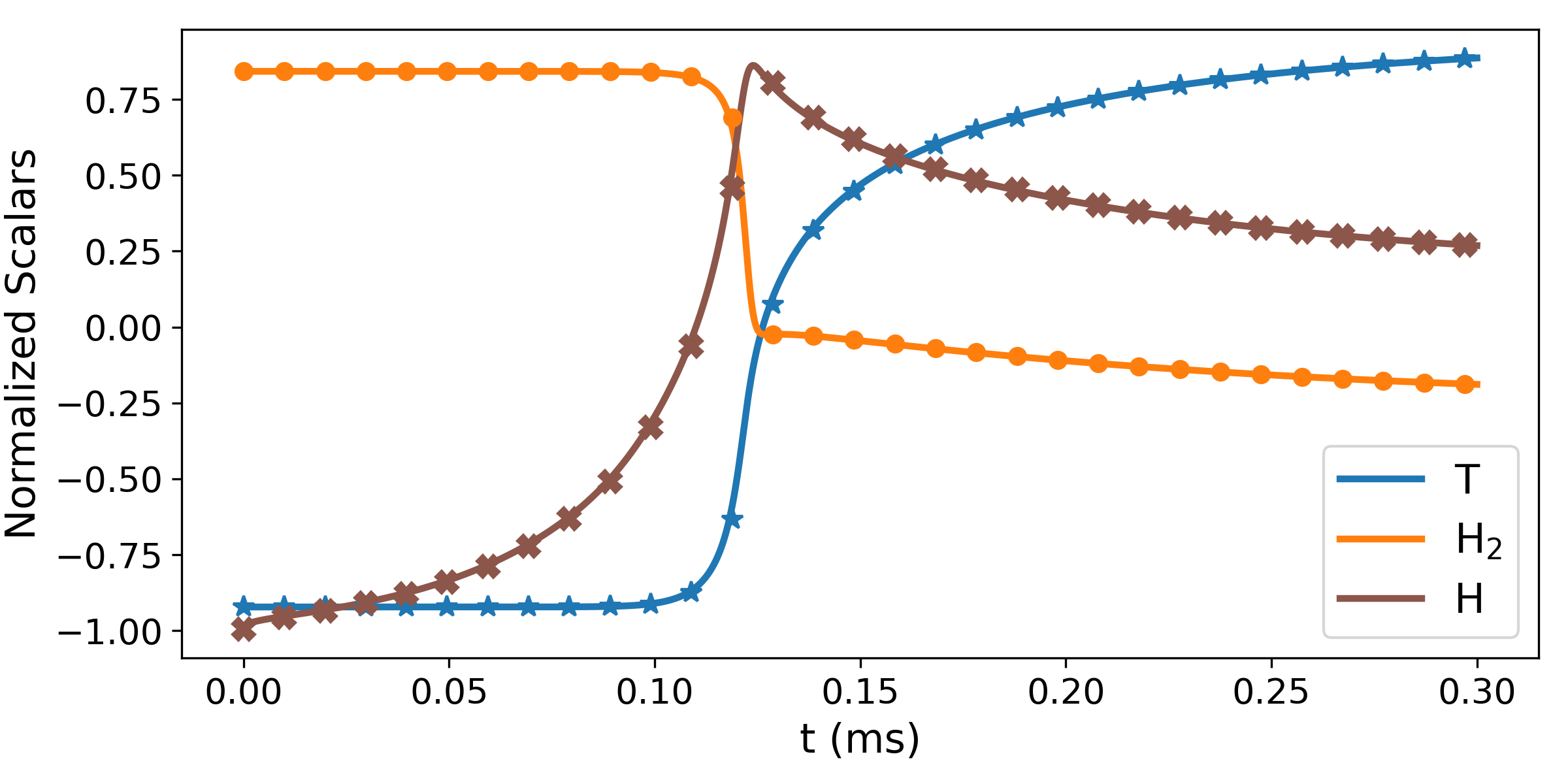} 
\caption{H$_2$,$\Phi_i = 1.0, T_i= 1065 K$}
\label{h2phi1t1065}
\end{subfigure}
\hfill
\begin{subfigure}[h]{0.48\textwidth}
\includegraphics[width= \textwidth]{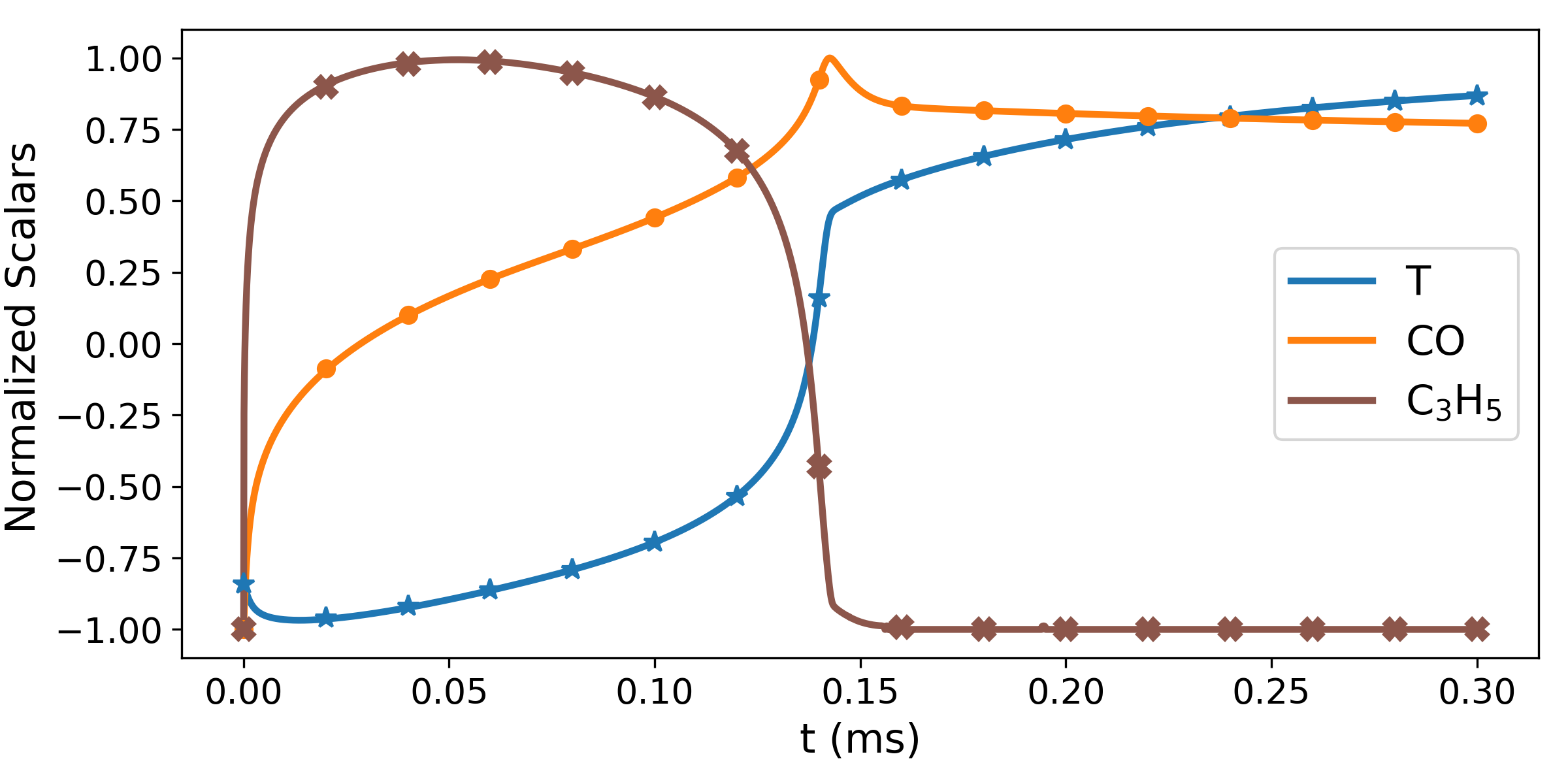}
\caption{High-Temperature n-dodecane $\Phi_i = 1.0, T_i= 1420 K$}
\label{hndti1420}
\end{subfigure}

\begin{subfigure}[h]{0.48\textwidth}
\includegraphics[width= \textwidth]{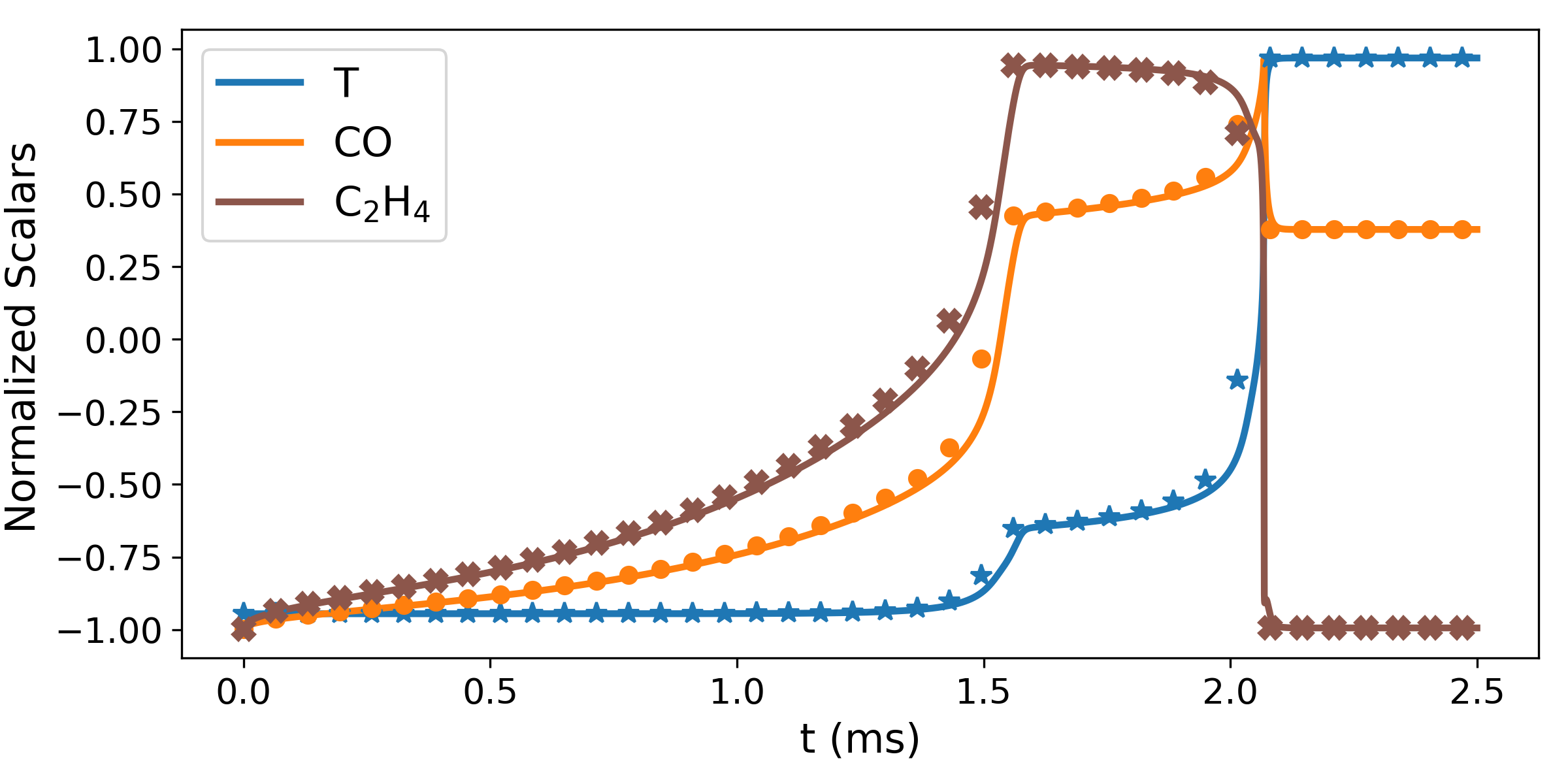}
\caption{Low-Temperature n-dodecane $\Phi_i = 1.0, T_i= 780 K$}
\label{lndti780}
\end{subfigure}
\hfill
\begin{subfigure}[h]{0.48\textwidth}
\includegraphics[width= \textwidth]{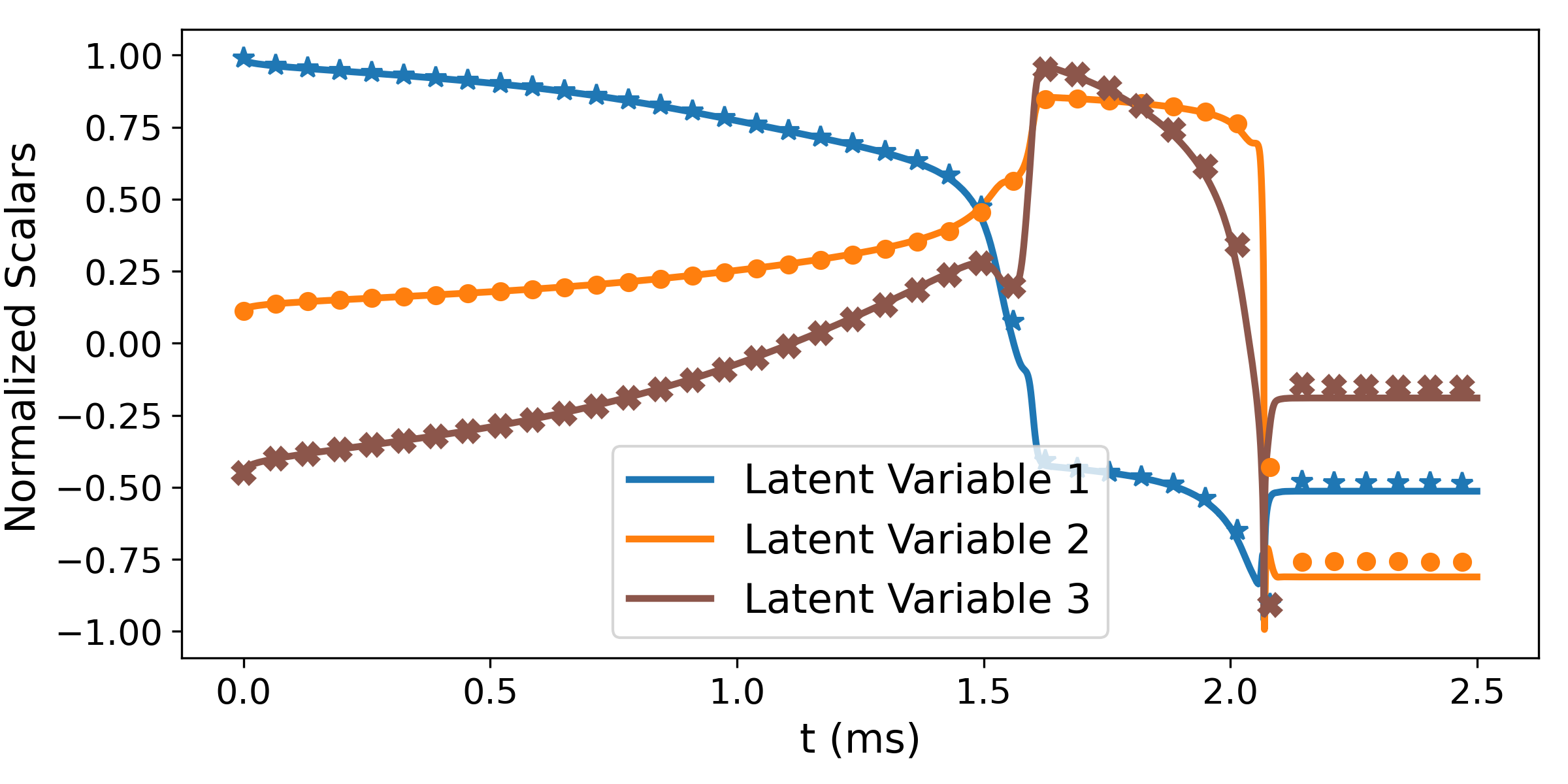}
\caption{Laten space Low-Temp n-dodecane $\Phi_i = 1.0, T_i= 780 K$}
\label{autoti780}
\end{subfigure}

\caption{Temporal evolution of normalized scalars. The solid line and symbols represent Cantera and React-DeepOnet solution respectively }
\label{tempevolution}
\end{figure}

Figure-~\ref{tempevolution} shows the temporal evolution of the normalized scalars. Plots -\ref{h2phi1t1065}, \ref{hndti1420} and \ref{lndti780} show the evolution of normalized temperature and species mass fractions for H$_2$, high-temperature n-dodecane and low-temperature n-dodecane oxidation respectively. Predictions from React-DeepOnet show an excellent agreement with the exact values from Cantera with minutely small mean absolute errors listed in Table- \ref{tab:table2}. From the simpler chemical reaction interaction of H$_2$ oxidation to the highly complex interaction of low-temperature n-dodecane oxidation, React-DeepOnet is able to accurately learn the combustion dynamics while reproducing the ignition delay times and equilibrium thermochemical values. The plot-\ref{autoti780} shows the temporal evolution of the normalized latent variables. Latent space React-DeepOnet accurately learns the complex low-temperature n-dodecane combustion dynamics in the reduced space with far smaller network size (less than half of low-temperature n-dodecane React-DeepOnet) as listed in Table-\ref{tab:table2}.

\begin{figure}[h!]
\centering 

\begin{subfigure}[h]{.48\textwidth}
\includegraphics[width= \textwidth]{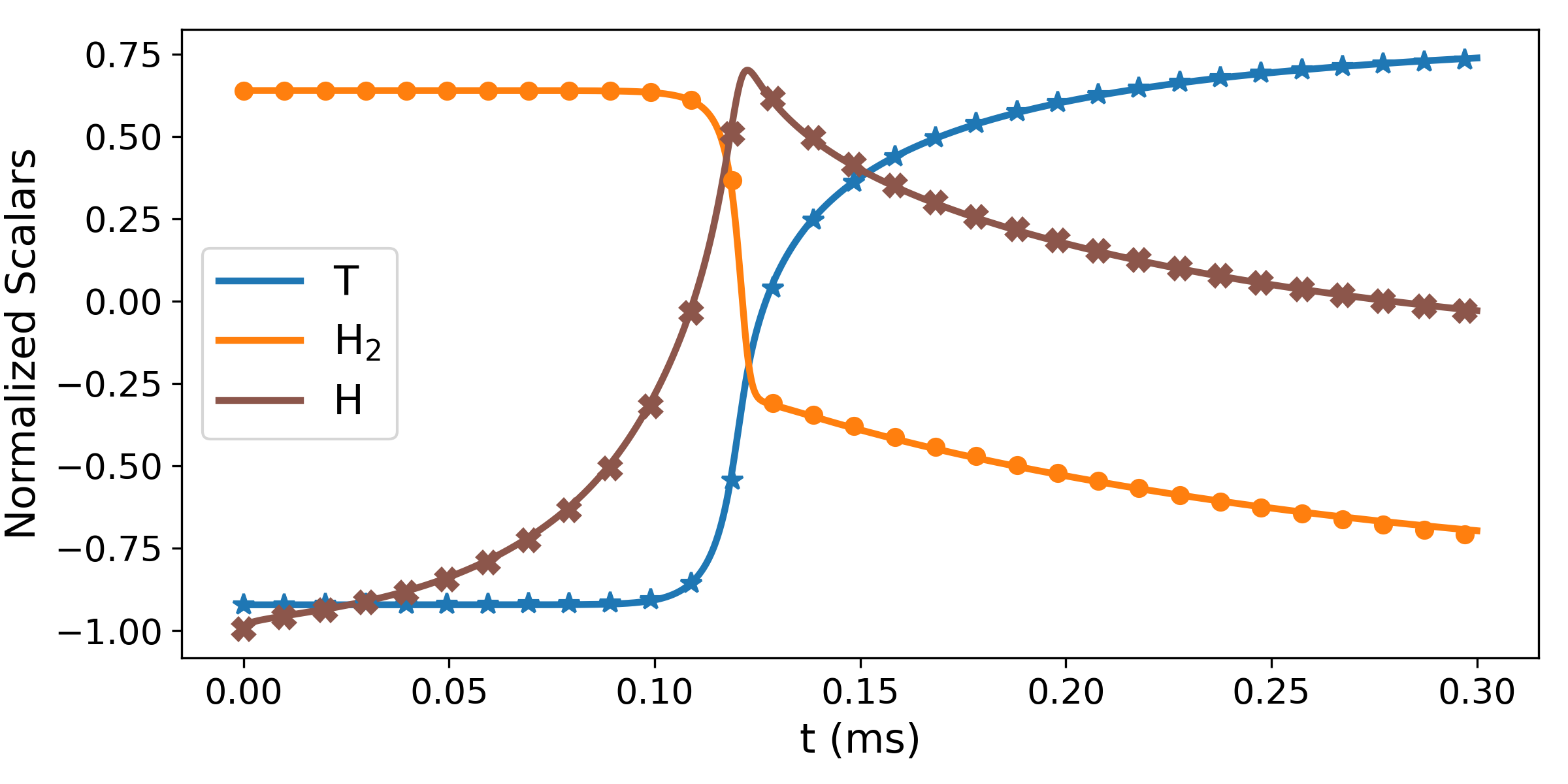}
\caption{H$_2$, $\Phi_i = 0.7, T_i= 1065 K$}
\label{h2phi07}
\end{subfigure}
\hfill
\begin{subfigure}[h]{.48\textwidth}
\includegraphics[width= \textwidth]{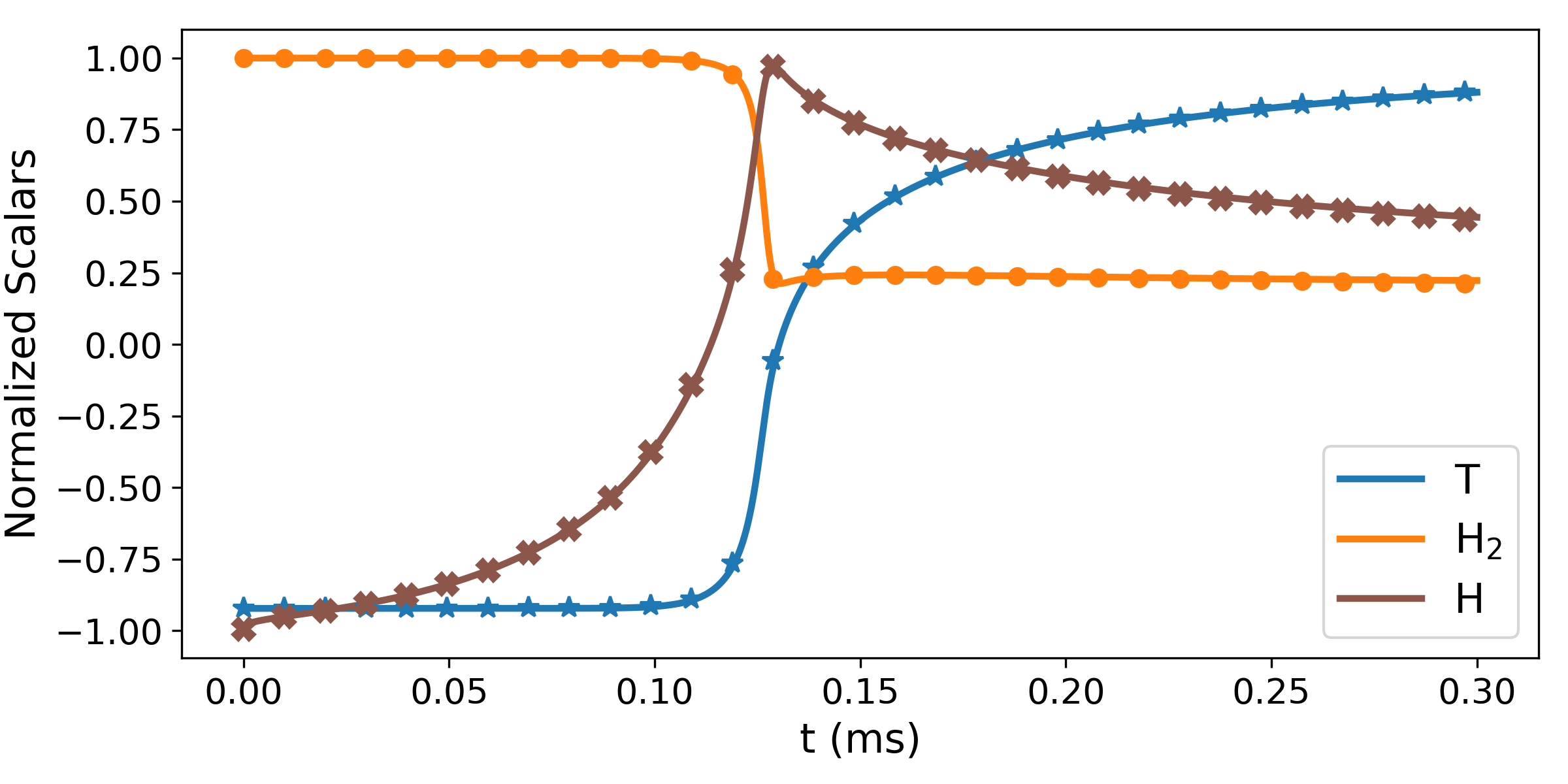}
\caption{H$_2$,$\Phi_i = 1.3, T_i= 1065 K$}
\label{h2phi13}
\end{subfigure}

\caption{Demonstrating extrapolation capabilities of React-DeepOnet: Predictions from React-DeepOnet are completely out of the training domain; The solid line and symbols represent Cantera and React-DeepOnet solution respectively.}
\label{h2extrpolate}
\end{figure}

Additionally, React-DeepOnet framework shows a great extrapolation capability. Figure- \ref{h2extrpolate} shows the predictions of React-DeepOnet at initial conditions which are completely out of the training domain. The predictions are in great agreement with the exact solution having mean absolute errors of $0.8 \%$, $0.6 \%$ for the respective figures. This states that the dynamics captured by React-DeepOnet are close to the physical laws and generalize well across unseen conditions.

\begin{figure}[h!]
\centering 

\begin{subfigure}[h]{.48\textwidth}
\includegraphics[width= \textwidth]{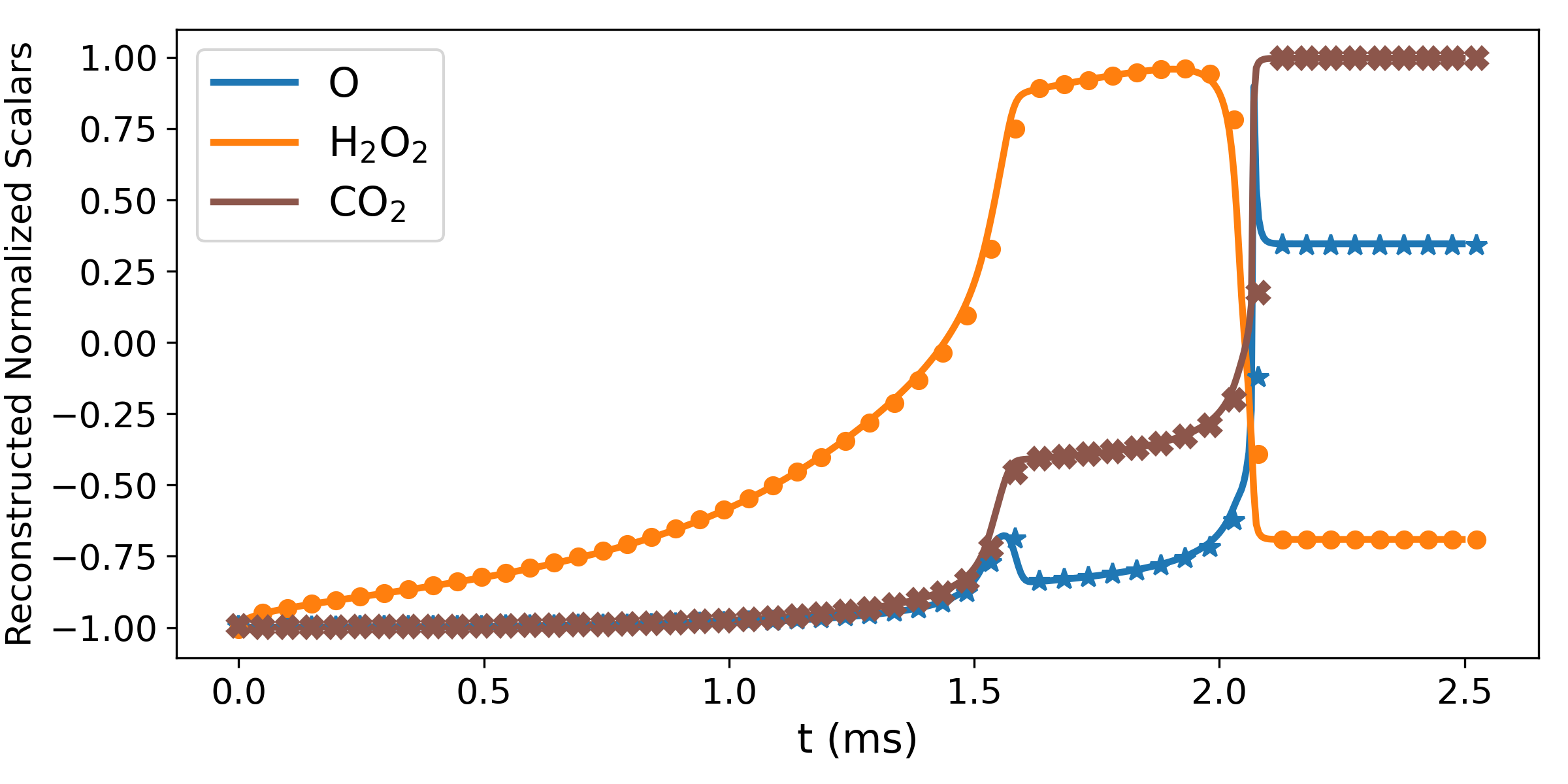}
\caption{Reconstructed O,H$_2$O$_2$,CO$_2$ at $T_i=$ 780 K}
\label{rnd1ti780}
\end{subfigure}
\hfill
\begin{subfigure}[h]{.48\textwidth}
\includegraphics[width= \textwidth]{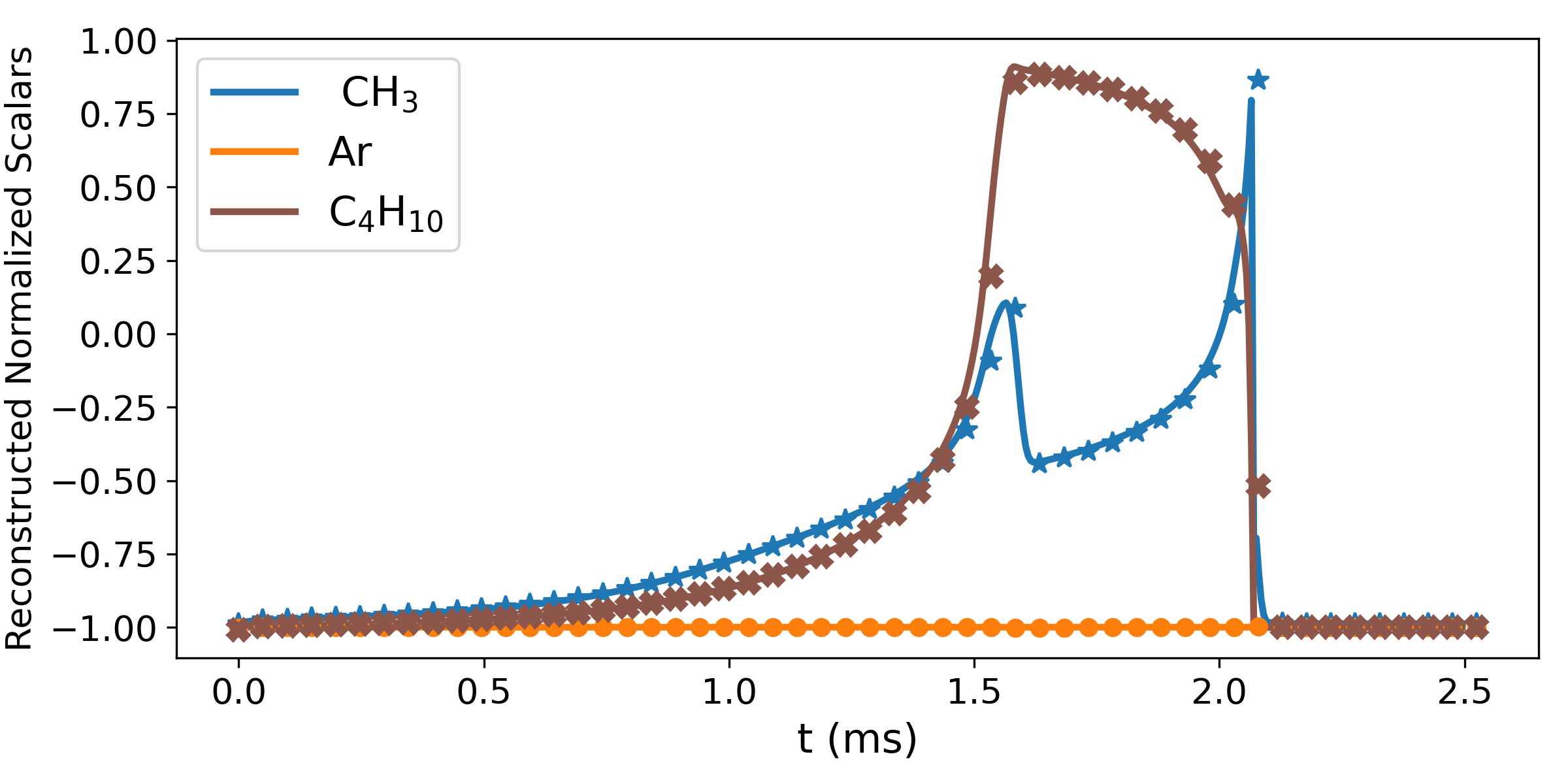}
\caption{Reconstructed CH$_3$,Ar,C$_4$H$_{10}$ at $T_i=$ 780 K}
\label{rnd2ti780}
\end{subfigure}

\caption{Reconstruction of remaining species with an ANN for low-temperature n-dodecane oxidation}
\label{rndphi1}
\end{figure}

Furthermore, leftover thermochemical scalars being correlated to the representative set can be non-linearly reconstructed through an ANN.
\[
\phi_j = f(\phi_i)
\]
where $i = [1, 2, \dots, m]$ and $j = [m+1, m+2, \dots, n]$. Figures-~\ref{rndphi1} shows the reconstruction of $Y_O, Y_{\rm{H_2O_2}}, Y_{\rm{CO_2}}$ and $Y_{\rm{CH_3}}, Y_{Ar}, Y_{\rm{C_4H_{10}}}$ respectively. Despite not being part of the solution vector, which is advanced through React-DeepOnet, these reconstructed scalars are almost identical to the exact values.

Table-~\ref{tab:table2} also summarizes the final solution integration time($t=T$) and CPU time taken to infer solutions from React-DeepOnet and Cantera ODE solver for each case. An impressive huge speed-up of $\mathcal{O}(3)$ is observed with React-DeepOnet for n-dodecane combustion. React-DeepOnet, corresponding to the simpler high-temperature chemistry, is smaller in size and around 12,000 times faster than the Cantera ODE solver. Low-temperature chemistry React-DeepOnet, despite its large size, is around 2,000 times faster than the Cantera solver. Furthermore, the latent space React-DeepOnet for low-temperature chemistry, owing to its smaller size, is around 4,000 times faster than Cantera. The reaction system for H$_2$ combustion is a lot simpler and as a result, the Cantera solver does not take a large time to integrate the solution. Despite that, React-DeepOnet for H$_2$ combustion is around 7 times faster.

\section{Conclusions}
A combustion chemistry acceleration scheme is proposed. It is based on DeepONets, which project solutions of thermochemical scalars' vectors over time increments. Two different computational efficiency enhancements are implemented, including the use of the ShiftNet and latent space dynamics identification using autoencoders. The approach is validated to predict hydrogen-air and n-dodecane low- and high-temperature oxidations. Validations of the approach show that it can predict the evolution of species and temperature over relevant integration times. Most importantly, the approach is enabling significant speedups in chemistry integration, which are more pronounced for more complex fuels.

The React-DeepOnet framework with its promising results of low generalization error and swift solution inference is a perfect replacement for traditional ODE solvers in combustion simulation. In the future,  a study on the extrapolation capability of React-DeepOnet for a larger reaction mechanism would make this framework comprehensive. The integration of React-DeepONet within the context of CFD simulations is a natural extension of this work and may help speed up such simulations. Finally, further validation with other complex fuels and the development of effective data sets for training React-DeepOnet will be an important focus of our group.

\section{Acknowledgements}
The code for React -DeepOnet framework is inspired by the base code of DeepOnets from \cite{longtime} 
(https://github.com/PredictiveIntelligenceLab/Long-time-Integration-PI-DeepONets).

\bibliographystyle{elsarticle-num} 
\bibliography{Comprehensive.bib}

\begin{thebibliography}{10}
\expandafter\ifx\csname url\endcsname\relax
  \def\url#1{\texttt{#1}}\fi
\expandafter\ifx\csname urlprefix\endcsname\relax\def\urlprefix{URL }\fi
\expandafter\ifx\csname href\endcsname\relax
  \def\href#1#2{#2} \def\path#1{#1}\fi

\bibitem{Pope1997}
S.~Pope, Computationally efficient implementation of combustion chemistry using
  in situ adaptive tabulation, Combust. Sci. Tech. 1 (1997) 41--63.

\bibitem{Tonse2003}
S.~Tonse, N.~Moriarty, M.~Frenklach, N.~Brown, Computational economy
  improvements in {{PRISM}}, Int. J. Chem. Kin. 35 (2003) 438--452.

\bibitem{liang2009}
L.~Liang, J.~Stevens, S.~Raman, J.~Farrell, The use of dynamic adaptive
  chemistry in combustion simulation of gasoline surrogate fuels, Combust.
  Flame 156 (2009) 1493--1502.

\bibitem{Continuo2011}
F.~Continuo, H.~Jeanmart, T.~Lucchini, G.~D'Errico, Coupling of in situ
  adaptive tabulation and dynamic adaptive chemistry: An effective method for
  solving combustion in engine simulations, Proc. Combust. Inst. 33 (2011)
  3057--3064.

\bibitem{Sun2017}
W.~Sun, Y.~Ju, Ta multi-timescale and correlated dynamic adaptive chemistry and
  transport {{(CO-DACT)}} method for computationally efficient modeling of jet
  fuel combustion with detailed chemistry and transport, Combust. Flame 184
  (2017) 297--311.

\bibitem{DAlessio2020}
G.~D'Alessio, A.~Cuoci, G.~Aversano, M.~Bracconi, A.~Stagni, A.~Parente,
  {Impact of the Partitioning Method on Multidimensional Adaptive-Chemistry
  Simulations}, Energies {13}~({10}) ({MAY} {2020}).

\bibitem{maas1992}
U.~Maas, S.~Pope, Simplifying chemical kinetics - intrinsic low-dimensional
  manifolds in composition space, Combust. Flame 88 (1992) 239--264.

\bibitem{Lam1994}
S.~Lam, D.~Goussis, {The CSP method for simplifying kinetics}, Int. J. Chem.
  Kin. 26 (1994) 461--486.

\bibitem{Christo1996a}
F.~Christo, A.~Masri, E.~Nebot, S.~Pope, An integrated pdf/neural network
  approach for simulating turbulent reacting systems, in: 26th Symp. (Int.) on
  Combustion, 1996, pp. 43--48.

\bibitem{Christo1996b}
F.~Christo, A.~Masri, E.~Nebot, {Artificial neural network implementation of
  chemistry with pdf simulation of H2/CO2 flames}, Combust. Flame 106 (1996)
  406--427.

\bibitem{Blasco1998}
J.~Blasco, N.~Fueyo, C.~Dopazo, J.~Ballester, Modelling the temporal evolution
  of a reduced combustion chemical system with an artificial neural network,
  Combust. Flame 113 (1998) 38--52.

\bibitem{Blasco2000}
J.~Blasco, N.~Fueyo, C.~Dopazo, J.~Chen, A self-organizing-map approach to
  chemistry representation in combustion applications, Combust. Theo. Model. 4
  (2000) 61--76.

\bibitem{Ranade2019c}
R.~Ranade, S.~Alqahtani, A.~Farooq, T.~Echekki, An ann based hybrid chemistry
  framework for complex fuels, Fuel 241 (2019) 625--636.

\bibitem{Ranade2019d}
R.~Ranade, S.~Alqahtani, A.~Farooq, T.~Echekki, An extended hybrid chemistry
  framework for complex hydrocarbon fuels, Fuel 251 (2019) 276--284.

\bibitem{Chen2000}
J.~Chen, J.~Blasco, N.~Fueyo, C.~Dopazo, An economical strategy for storage of
  chemical kinetics: Fitting in situ adaptive tabulation with artificial neural
  networks, Proc. Combust. Inst. 28 (2000) 115--121.

\bibitem{Wan2020}
K.~Wan, C.~Barnaud, L.~Vervisch, P.~Domingo, {Chemistry reduction using machine
  learning trained from non-premixed micro-mixing modeling: Application to DNS
  of a syngas turbulent oxy-flame with side-wall effects}, {Combust. Flame}
  {220} ({2020}) {119--129}.

\bibitem{Wan2021}
K.~Wan, C.~Barnaud, L.~Vervisch, P.~Domingo, {Machine learning for detailed
  chemistry reduction in DNS of a syngas turbulent oxy-flame with side-wall
  effects}, Proc. Combust. Inst. {38} ({2021}) {2825--2833}.

\bibitem{Alqahtani2021}
S.~Alqahtani, T.~Echekki, A data-based hybrid model for complex fuel chemistry
  acceleration at high temperatures, Combust. Flame 223 (2021) 142--152.

\bibitem{Echekki2023}
T.~Echekki, A.~Farooq, M.~Ihme, S.~M. Sarathy, Machine Learning for Combustion
  Chemistry, Springer International Publishing, Cham, 2023, pp. 117--147.

\bibitem{ChemNODE}
O.~Owoyele, P.~Pal, {ChemNODE: A neural ordinary differential equations
  framework for efficient chemical kinetic solvers}, Energy AI {7} ({2022}).

\bibitem{Zhang2021AIAA}
T.~Zhang, Y.~Zhang, W.~E, Y.~Ju, {DLODE}: A deep learning-based ode solver for
  chemical kinetics~(AIAA Paper 2021-1139) (2021).

\bibitem{Galassi2022}
R.~Galassi, P.~Ciottoli, M.~Valorani, H.~Im, An adaptive time-integration
  scheme for stiff chemistry based on computational singular perturbation and
  artificial neural networks, J. Comput. Phys. (2022).

\bibitem{deeponet_orig}
L.~Lu, P.~Jin, G.~Pang, Z.~Zhang, G.~E. Karniadakis, Learning nonlinear
  operators via deeponet based on the universal approximation theorem of
  operators, Nature Machine Intelligence 3~(3) (2021) 218--229.

\bibitem{universal_operator}
T.~Chen, H.~Chen, Universal approximation to nonlinear operators by neural
  networks with arbitrary activation functions and its application to dynamical
  systems, IEEE Transactions on Neural Networks 6~(4) (1995) 911--917.

\bibitem{longtime}
S.~Wang, P.~Perdikaris, Long-time integration of parametric evolution equations
  with physics-informed deeponets (06 2021).

\bibitem{Mao2021}
Z.~Mao, L.~Lu, O.~Marxen, T.~A. Zaki, G.~E. Karniadakis, Deepm\&mnet for
  hypersonics: Predicting the coupled flow and finite-rate chemistry behind a
  normal shock using neural-network approximation of operators, J. Comput.
  Phys. 447 (2021).

\bibitem{WangWangPerdikaris2021}
S.~Wang, H.~Wang, P.~Perdikaris, Learning the solution operator of parametric
  partial differential equations with physics-informed deeponets, Science
  Advances 7~(40) (2021) eabi8605.

\bibitem{Venturi2023}
S.~Venturi, T.~Casey, Svd perspectives for augmenting deeponet flexibility and
  interpretability, Comput. Meth. Appl. Mech. Engg. 403 (2023) 115718.

\bibitem{hadorn2022shift}
P.~S. Hadorn, Shift-deeponet: Extending deep operator networks for
  discontinuous output functions, ETH Zurich, Seminar for Applied Mathematics,
  2022.

\bibitem{Liu2022DeepPropNetA}
L.~Liu, W.~Cai, Deeppropnet - a recursive deep propagator neural network for
  learning evolution pde operators, ArXiv abs/2202.13429 (2022).

\bibitem{cantera}
D.~G. Goodwin, R.~L. Speth, H.~K. Moffat, B.~W. Weber, Cantera: An
  object-oriented software toolkit for chemical kinetics, thermodynamics, and
  transport processes, \url{https://www.cantera.org}, version 2.5.1 (2021).

\bibitem{Owoyele2017}
O.~Owoyele, T.~Echekki, Toward computationally efficient combustion
  \uppercase{DNS} with complex fuels via principal component transport,
  Combust. Theo. Model. 21 (2017) 770--798.

\bibitem{mirgol2015}
H.~Mirgolbabaei, T.~Echekki, N.~Smaoui, A nonlinear principal component
  analysis approach for turbulent combustion composition space, Int. J.
  Hydrogen Energy 39 (2014) 4622--4633.

\bibitem{brunton2019}
K.~Champion, B.~Lusch, J.~N. Kutz, S.~L. Brunton, Data-driven discovery of
  coordinates and governing equations, Proceedings of the National Academy of
  Sciences 116~(45) (2019) 22445--22451.

\bibitem{carlberg2020}
K.~Lee, K.~T. Carlberg, Model reduction of dynamical systems on nonlinear
  manifolds using deep convolutional autoencoders, J. Comput. Phys. 404 (2020)
  108973.

\bibitem{zhang2022}
H.~Dikeman, H.~Zhang, S.~Yang, Stiffness-reduced neural ode models for
  data-driven reduced-order modeling of combustion chemical kinetics, 2022.
\newblock \href {https://doi.org/10.2514/6.2022-0226}
  {\path{doi:10.2514/6.2022-0226}}.

\bibitem{grassi2022reducing}
T.~Grassi, F.~Nauman, J.~Ramsey, S.~Bovino, G.~Picogna, B.~Ercolano, Reducing
  the complexity of chemical networks via interpretable autoencoders, Astronomy
  \& Astrophysics 668 (2022) A139.

\bibitem{fries2022}
W.~D. Fries, X.~He, Y.~Choi, {LaSDI: Parametric Latent Space Dynamics
  Identification}, Comp. Meth. Appl. Mech. Engg 399 (2022) 115436.

\bibitem{wang2010high}
H.~Wang, E.~Dames, B.~Sirjean, D.~Sheen, R.~Tangko, A.~Violi, J.~Lai,
  F.~Egolfopoulos, D.~Davidson, R.~Hanson, et~al., A high-temperature chemical
  kinetic model of n-alkane (up to n-dodecane), cyclohexane, and methyl-,
  ethyl-, n-propyl and n-butyl-cyclohexane oxidation at high temperatures,
  JetSurF version 2~(2) (2010) 19.

\bibitem{frosting2018}
R.~Frostig, M.~Johnson, C.~Leary,
  \href{https://mlsys.org/Conferences/doc/2018/146.pdf}{Compiling machine
  learning programs via high-level tracing}, 2018.
\newline\urlprefix\url{https://mlsys.org/Conferences/doc/2018/146.pdf}

\bibitem{kingma2014}
D.~Kingma, J.~Ba, Adam: A method for stochastic optimization, International
  Conference on Learning Representations (12 2014).

\end{thebibliography}
\end{document}